\documentclass{llncs}

\usepackage{natbib}
\usepackage{graphicx}

\usepackage{fancyhdr}
\usepackage{lastpage}

\usepackage{booktabs}
\usepackage{tablefootnote}
 
\begin{document}

\title{A short review and primer on the use of human voice in human computer interaction applications}
\author{Kristian Lukander}
\institute{Quantified Employee unit, Finnish Institute of Occupational Health,\\
\email{kristian.lukander@ttl.fi},\\
POBox 40, Helsinki, 00250, Finland}

\maketitle              

\begin{abstract}
The application of psychophysiologicy in human-computer interaction is a growing field with significant potential for future smart personalised systems. Working in this emerging field requires comprehension of an array of physiological signals and analysis techniques. 

Human speech affords, alongside linguistic content, rich information in the intonation, voice quality, prosody, and rhythmic variation of utterances, allowing listeners to recognise numerous distinct emotional states in the speaker. Several types of factors affect speech, ranging from emotions to cognitive load and pathological conditions, providing a promising non-intrusive source for online understanding of context and psychophysiological state.

This paper aims to serve as a primer for the novice, enabling rapid familiarisation with the latest core concepts. We put special emphasis on everyday human-computer interface applications to distinguish from the more common clinical or sports uses of psychophysiology.

This paper is an extract from a comprehensive review of the entire field of ambulatory psychophysiology, including 12 similar chapters, plus application guidelines and systematic review. Thus any citation should be made using the following reference:

{\parshape 1 2cm \dimexpr\linewidth-1cm\relax
B. Cowley, M. Filetti, K. Lukander, J. Torniainen, A. Henelius, L. Ahonen, O. Barral, I. Kosunen, T. Valtonen, M. Huotilainen, N. Ravaja, G. Jacucci. \textit{The Psychophysiology Primer: a guide to methods and a broad review with a focus on human-computer interaction}. Foundations and Trends in Human-Computer Interaction, vol. 9, no. 3-4, pp. 150--307, 2016.
\par}


\end{abstract}

Human voice production originates at the larynx, where air pressure from the lungs causes vibration of the vocal folds, thereby generating a complex but patterned sound source composed of a fundamental frequency and multiple harmonics. This signal is then filtered through the vocal tract airways (oral and nasal cavities). This vocal apparatus produces a complex interactive system capable of generating a wide variety of sounds \citep{ghazanfar2008}. 

Humans have evolved a nonverbal communication system in which, alongside linguistic content, speech carries rich information in the intonation, voice quality, prosody, and rhythmic variation of utterances, allowing listeners to recognise numerous distinct emotional states in the speaker. Several types of factors, from emotions to cognitive load and pathological conditions, affect the functioning of the larynx, or `voice box', whereby the internal state of the speaker causes tightening/relaxation of the vocal folds, which modulates the acoustic and rhythmic components of speech.

Now that computing systems are starting to listen actively to people (as with Google's `Voice', Microsoft's `Cortana', and Apple's `Siri'), human speech provides a promising source for online understanding of context and psychophysiological state with measurement that remains minimally intrusive. 

\section{Background}
Evaluating stress, affect, and mood on the basis of the human voice is not a new notion. Much of the literature on the effects of psychophysiological states on the acoustics of speech production has its roots in the 1970s, in work on stress and lie detectors in interrogation, military, and aerospace settings. Technologies used in these settings -- namely, voice stress analysis, voice risk analysis, and layered voice analysis -- use the recorded features, including`micro tremors', in a person's voice to construct a scorable `voice gram', which is then evaluated by a specialist. While these approaches have been utilised in courtrooms and operator monitoring in demanding work tasks, their reliability has been disputed -- for these methods, the most convincing results have been obtained in conditions of extreme stress, such as under threat of injury or great operational risk. The recording environments in such research has possessed heterogeneous acoustic characteristics, and the results and metrics from the relevant studies cannot be cross-evaluated reliably \citep{hopkins2005, harnsberger2009}. 

However, even everyday user interaction situations appear to elicit strong enough emotional and stress responses to produce systematic, detectable changes in voice parameters. In the interactive setting of HCI specifically, the psychophysiological states recognised in the literature as having an effect on speech production are cognitive workload, or `stress' \citep{lively1993}; physical stress \citep{godin2015}, and various emotional states \citep{elayadi2011}. 

\section{Methods}
Changes in human voice production can be measured with a microphone, in combination with the use of mathematical models to associate acoustic changes with the functioning of the larynx, or through electroglottography, in which the system uses two pairs of electrodes (one pair on either side of the subject's throat) to measure the variations over time in the degree of contact of the vocal folds during voice production \citep{kania2006}. When microphones are employed, the approach typically involves inverse-filtering microphone recordings to model the waveform of the glottal airflow pulses, which, in turn, reflect the movements of the vocal folds \citep{alku2011}. There is a downside, however: microphone recordings vary greatly in quality and with respect to noise parameters related to microphone type, environmental factors, and distance from the speaker. Guidelines for selection of microphones suitable for human voice research are provided in a summary by \citet{svec2010}. 

Several derived low-level features and combinations thereof have been suggested to be correlated with variations in internal states. Table \ref{acousticfeatures}, compiled on the basis of work by \citet{ververidis2006} and \citet{scherer2003}, presents the effects of fundamental emotional state on selected features. Typical features derived from speech signals include loudness, fundamental frequency, word and utterance rate (speed), jitter, zero-crossing rate, and frequency ratios. While acoustic features of speech such as pitch, timing, voice quality, and articulation have been shown to correlate highly with underlying emotional activation (low--high), there is no agreement as to how these dimensions correlate with the valence of the emotion \citep{elayadi2011}. Instead of direct comparisons involving individual features or combinations of them, modern approaches tend to use machine learning methods to improve detection rates \citep{zhou2001}. 


\begin{table}[ht]
\centering
\caption{Effects of certain fundamental emotional states on selected acoustic features, compiled from work by \citet{scherer2003} and \citet{ververidis2006}. }
\label{acousticfeatures}
\begin{tabular}{llccccccc}
 & & Stress & Anger/rage & Fear/panic & Disgust & Sadness & Joy/elation & Boredom \\
 \toprule \\
Intensity & Mean & $\nearrow$ & $\nearrow$ & $\nearrow$ & $\searrow$ & $\searrow$ & $\nearrow$ & \\
 & Range & & $\nearrow$ & $\nearrow$ & & $\searrow$ & $\nearrow$ & \\
F0\tablefootnote{ Fundamental frequency} & floor/mean & $\nearrow$ & $\nearrow$ & $\nearrow$ & $\searrow$ & $\searrow$ & $\nearrow$ & \\
 & variability & & $\nearrow$ & & & $\searrow$ & $\nearrow$ & $\searrow$ \\ 
 & range & & $\nearrow$ & $\nearrow$ & $\nearrow$ $\searrow$\tablefootnote{ According to \citet{ververidis2006}, the F0 range among males increases.} & $\searrow$ & $\nearrow$ & 
 $\searrow$ \\
\multicolumn{2}{l}{Sentence contours} & & $\searrow$ & $\nearrow$ & &--\tablefootnote{ Studies do not agree.} & $\searrow$ & \\ 
\multicolumn{2}{l}{High-frequency energy} & & $\nearrow$ & $\nearrow$ & & $\searrow$ & $\nearrow$ & \\
\multicolumn{2}{l}{Speech and articulation rate} & & $\nearrow$ $\searrow$\tablefootnote{ According to \citet{scherer2003}, the articulation rate among males decreases.} & $\nearrow$ & $\searrow$ & $\nearrow$ $\searrow$\tablefootnote{ According to \citet{ververidis2006}, the articulation rate among males increases.} & $\nearrow$ & $\searrow$ \\
\multicolumn{2}{l}{Duration} & & $\searrow$ & $\searrow$ & & $\nearrow$ & $\searrow$ &
\end{tabular}
\end{table}


\section{Applications}

While variations between one speaker and the next require calibration and baselines, online calculation of acoustic parameters and spectral measures is relatively easy and robust. In one recent innovation, a set of open-source toolboxes for automated feature extraction and voice analysis has been developed. These include openSMILE -- the Munich Versatile and Fast Open-Source Audio Feature Extractor \citep{eyben2010} -- and AMMON (Affective and Mental Health Monitor) \citep{chang2011}.

A considerable amount of the recent research in this field has been presented in connection with the Interspeech computational paralinguistic challenges \citep{schuller2015}. Since 2009, these challenges have called for methods of evaluating speech for the detection of age, gender, affect and emotional parameters, personality, likeability, pathologies and diseases, social signals, signs of conflict, cognitive load, and physical demands. For an extensive list of examples of computational solutions for the paralinguistic detection of cognitive and emotional states, the reader is directed to the Interspeech repository available online at \url{http://compare.openaudio.eu/}. 

\section{Conclusion}
For the interactive setting that is the focus of our attention, analysing the acoustics of speech production offers a non-intrusive online metric for gauging the internal state of the user. There is considerable potential on account of the pervasiveness and unobtrusive nature of the method. The classification performance of automated solutions is beginning to reach an acceptable level of sensitivity and reliability, at least upon user-specific calibration to accommodate the effects of differences in languages and dialects, individual-to-individual differences in speech production, and variations in stress and affective responses. 

Another source of motivation for significant improvements in both recognising and producing natural paralinguistic cues associated with empathetic responses arises from future needs related to affective computing, robotics, and artificial conversation partners in general. These technologies necessitate naturalistic input and output down to the smallest detail in order to escape the `uncanny valley' \citep{mori2012}. 

\bibliographystyle{plainnat}
\bibliography{preprint_speech}

\begin{thebibliography}{16}
\providecommand{\natexlab}[1]{#1}
\providecommand{\url}[1]{\texttt{#1}}
\expandafter\ifx\csname urlstyle\endcsname\relax
  \providecommand{\doi}[1]{doi: #1}\else
  \providecommand{\doi}{doi: \begingroup \urlstyle{rm}\Url}\fi

\bibitem[Alku(2011)]{alku2011}
Paavo Alku.
\newblock Glottal inverse filtering analysis of human voice production--a
  review of estimation and parameterization methods of the glottal excitation
  and their applications.
\newblock \emph{Sadhana}, 36\penalty0 (5):\penalty0 623--650, 2011.

\bibitem[Chang et~al.(2011)Chang, Fisher, Canny, and Hartmann]{chang2011}
Keng-hao Chang, Drew Fisher, John Canny, and Bj{\"o}rn Hartmann.
\newblock How's my mood and stress?: An efficient speech analysis library for
  unobtrusive monitoring on mobile phones.
\newblock In \emph{Proceedings of the 6th International Conference on Body Area
  Networks}, pages 71--77. ICST (Institute for Computer Sciences,
  Social-Informatics and Telecommunications Engineering), 2011.

\bibitem[El~Ayadi et~al.(2011)El~Ayadi, Kamel, and Karray]{elayadi2011}
Moataz El~Ayadi, Mohamed~S Kamel, and Fakhri Karray.
\newblock Survey on speech emotion recognition: Features, classification
  schemes, and databases.
\newblock \emph{Pattern Recognition}, 44\penalty0 (3):\penalty0 572--587, 2011.

\bibitem[Eyben et~al.(2010)Eyben, W{\"o}llmer, and Schuller]{eyben2010}
Florian Eyben, Martin W{\"o}llmer, and Bj{\"o}rn Schuller.
\newblock Opensmile: The munich versatile and fast open-source audio feature
  extractor.
\newblock In \emph{Proceedings of the international conference on Multimedia},
  pages 1459--1462. ACM, 2010.

\bibitem[Ghazanfar and Rendall(2008)]{ghazanfar2008}
Asif~A Ghazanfar and Drew Rendall.
\newblock Evolution of human vocal production.
\newblock \emph{Current Biology}, 18\penalty0 (11):\penalty0 R457--R460, 2008.

\bibitem[Godin and Hansen(2015)]{godin2015}
Keith~W Godin and John~HL Hansen.
\newblock Physical task stress and speaker variability in voice quality.
\newblock \emph{EURASIP Journal on Audio, Speech, and Music Processing},
  2015\penalty0 (1):\penalty0 1--13, 2015.

\bibitem[Harnsberger et~al.(2009)Harnsberger, Hollien, Martin, and
  Hollien]{harnsberger2009}
James~D Harnsberger, Harry Hollien, Camilo~A Martin, and Kevin~A Hollien.
\newblock Stress and deception in speech: Evaluating layered voice analysis*.
\newblock \emph{Journal of forensic sciences}, 54\penalty0 (3):\penalty0
  642--650, 2009.

\bibitem[Hopkins et~al.(2005)Hopkins, Ratley, Benincasa, and
  Grieco]{hopkins2005}
Clifford~S Hopkins, Roy~J Ratley, Daniel~S Benincasa, and John~J Grieco.
\newblock Evaluation of voice stress analysis technology.
\newblock In \emph{System Sciences, 2005. HICSS'05. Proceedings of the 38th
  Annual Hawaii International Conference on}, pages 20b--20b. IEEE, 2005.

\bibitem[Kania et~al.(2006)Kania, Hartl, Hans, Maeda, Vaissiere, and
  Brasnu]{kania2006}
Romain~E Kania, Dana~M Hartl, St{\'e}phane Hans, Shinji Maeda, Jacqueline
  Vaissiere, and Daniel~F Brasnu.
\newblock Fundamental frequency histograms measured by electroglottography
  during speech: A pilot study for standardization.
\newblock \emph{Journal of voice}, 20\penalty0 (1):\penalty0 18--24, 2006.

\bibitem[Lively et~al.(1993)Lively, Pisoni, Van~Summers, and
  Bernacki]{lively1993}
Scott~E Lively, David~B Pisoni, W~Van~Summers, and Robert~H Bernacki.
\newblock Effects of cognitive workload on speech production: Acoustic analyses
  and perceptual consequences.
\newblock \emph{The Journal of the Acoustical Society of America}, 93\penalty0
  (5):\penalty0 2962--2973, 1993.

\bibitem[Mori et~al.(2012)Mori, MacDorman, and Kageki]{mori2012}
Masahiro Mori, Karl~F MacDorman, and Norri Kageki.
\newblock The uncanny valley [from the field].
\newblock \emph{Robotics \& Automation Magazine, IEEE}, 19\penalty0
  (2):\penalty0 98--100, 2012.

\bibitem[Scherer(2003)]{scherer2003}
Klaus~R Scherer.
\newblock Vocal communication of emotion: A review of research paradigms.
\newblock \emph{Speech communication}, 40\penalty0 (1):\penalty0 227--256,
  2003.

\bibitem[Schuller et~al.(2015)Schuller, Steidl, and A]{schuller2015}
B~Schuller, S~Steidl, and Batliner A.
\newblock Repository of the interspeech computational paralinguistics challenge
  (compare) series.
\newblock \url{http://www.compare.openaudio.eu/}, 2015.
\newblock Accessed:2016-01-15.

\bibitem[Svec and Granqvist(2010)]{svec2010}
Jan~G Svec and Svante Granqvist.
\newblock Guidelines for selecting microphones for human voice production
  research.
\newblock \emph{American Journal of Speech-Language Pathology}, 19\penalty0
  (4):\penalty0 356--368, 2010.

\bibitem[Ververidis and Kotropoulos(2006)]{ververidis2006}
Dimitrios Ververidis and Constantine Kotropoulos.
\newblock Emotional speech recognition: Resources, features, and methods.
\newblock \emph{Speech communication}, 48\penalty0 (9):\penalty0 1162--1181,
  2006.

\bibitem[Zhou et~al.(2001)Zhou, Hansen, and Kaiser]{zhou2001}
Guojun Zhou, John~HL Hansen, and James~F Kaiser.
\newblock Nonlinear feature based classification of speech under stress.
\newblock \emph{Speech and Audio Processing, IEEE Transactions on}, 9\penalty0
  (3):\penalty0 201--216, 2001.

\end{thebibliography}

\end{document}